\documentclass[aps,prl,reprint,amsmath,amssymb,nofootinbib,floatfix,superscriptaddress]{revtex4-2}
\usepackage{graphicx}
\usepackage{dcolumn}
\usepackage{bm}
\usepackage{float}
\usepackage{dsfont} 
\usepackage{braket}
\usepackage{xcolor}
\usepackage{mathbbol}
\usepackage{tabularx}
\usepackage{booktabs}
\usepackage{orcidlink}
\usepackage{hyperref}
\hypersetup{colorlinks=true,citecolor= blue}

\usepackage[english]{babel}
\usepackage{enumerate}
\usepackage{multibib}
\usepackage{enumitem}
\newcites{New}{References}

\newcommand{\beginsupplement}{
  \setcounter{table}{0}
  \renewcommand{\thetable}{S\arabic{table}}%
  \setcounter{figure}{0}
  \renewcommand{\thefigure}{S\arabic{figure}}%
  \setcounter{equation}{0}
  \renewcommand{\theequation}{S\arabic{equation}}%
 } 
\newcommand{\beq}{\begin{equation}}
	\newcommand{\eeq}{\end{equation}}
\newcommand{\beqa}{\begin{eqnarray}}
                     \newcommand{\eeqa}{\end{eqnarray}}

\def\Xint#1{\mathchoice
       {\XXint\displaystyle\textstyle{#1}}%
       {\XXint\textstyle\scriptstyle{#1}}%
       {\XXint\scriptstyle\scriptscriptstyle{#1}}%
       {\XXint\scriptscriptstyle\scriptscriptstyle{#1}}%
       \!\int}
    \def\XXint#1#2#3{{\setbox0=\hbox{$#1{#2#3}{\int}$}
         \vcenter{\hbox{$#2#3$}}\kern-.5\wd0}}
    
    \def\dashint{\Xint-}
    
\begin{document}

\title{Exact emulation of few-body systems at low cost}

\author{Sven Heihoff\orcidlink{0009-0000-3641-0640}}
\affiliation{Ruhr-Universit\"at Bochum, Fakult\"at f\"ur Physik und Astronomie, Institut f\"ur Theoretische Physik II, D-44780 Bochum, Germany}

\author{Arseniy A.~Filin\orcidlink{0000-0002-7603-451X}}
\affiliation{Ruhr-Universit\"at Bochum, Fakult\"at f\"ur Physik und Astronomie, Institut f\"ur Theoretische Physik II, D-44780 Bochum, Germany}

\author{Evgeny Epelbaum\orcidlink{0000-0002-7613-0210}}
\affiliation{Ruhr-Universit\"at Bochum, Fakult\"at f\"ur Physik und Astronomie, Institut f\"ur Theoretische Physik II, D-44780 Bochum, Germany}

\begin{abstract}
Effective field theories have established themselves as key pillars of modern nuclear physics. They enable a quantitative understanding of the strong nuclear force, provided low-energy constants that parametrize short-distance physics can be determined from experimental data.
This, however, often becomes prohibitively expensive due to a significant computational cost of solving the $A$-body problem. The computational challenge is particularly severe for three-body forces, which are at the frontier of nuclear and atomic physics and play an important role in the equation of state of neutron stars. Here we prove that for a parametric low-rank update of a Hamiltonian, the $A$-body problem at a fixed energy {\it exactly} reduces to a low-dimensional matrix equation regardless of the size of the Hilbert space. As a proof-of-principle, we present {\it exact} and computationally cheap snapshot-based emulators for few-body scattering and bound states. Unlike alternatives, our emulators can be used far away from the snapshot region without loss of precision and yield accurate results for parameter values not accessible using conventional solution techniques. Our approach is not restricted by the interaction type, number of particles, and methods for generating snapshots and can be applied to mitigate the computational burden of the $A$-body problem to a broad class of problems in nuclear, atomic, and molecular physics.
\end{abstract}

\maketitle

{\it Introduction.---}
The quest for developing accurate and precise nuclear interactions in chiral effective field theory (EFT) \cite{Epelbaum:2008ga} puts new requirements on computational few- and many-body physics. In particular, fixing unknown low-energy constants (LECs), whose number grows with the expansion order, from experimental data requires performing numerous calculations of bound-state and scattering observables by solving the quantum $A$-body problem, which is computationally expensive. This is particularly pressing for three-nucleon (3N) forces \cite{Endo:2024cbz}, whose contributions at fifth chiral order involve a large number of  LECs \cite{Girlanda:2011fh,Huesmann:2026khj}. The use of Bayesian methods for inferring LECs from data and uncertainty quantification \cite{Wesolowski:2015fqa,Sun:2024iht} make the problem even more relevant. 

In response to this challenge, efforts are being made to develop emulators for a computationally efficient implementation of parametric updates of the Hamiltonian in $A$-body systems, see Refs.~\cite{Witala:2021xqm,Wesolowski:2021cni,Bonilla:2022rph,Garcia:2023slj,Sun:2024iht,Armstrong:2025tza,Gnech:2025lbg} for examples. Most of these emulators utilize reduced basis methods (RBM) by first solving the system for several sets of parameter values using conventional techniques. The resulting solutions, called snapshots, are then used to approximate the solution for other parameter values. The most prominent representative of such snapshot-based emulators relies on the eigenvector continuation (EC) method \cite{Frame:2017fah}, see Ref.~\cite{Duguet:2023wuh} for a review. The EC and other types of emulators have demonstrated their ability to trade accuracy for speed. 

In this Letter, we prove that under certain conditions (fixed energy and low-rank affine-type parametric update), one can drastically reduce the dimensionality of the $A$-body problem without accuracy loss. This allows us to build low-dimensional snapshot-based RBM emulators both for scattering and bound states, which reproduce the solution of the full problem in a {\it numerically exact} way. Our method opens up new perspectives in chiral EFT and in nuclear physics in general and can be used in other areas. 


{\it Two-body scattering.---} We start with the Lippmann-Schwinger-like equation (LSE) for the T-matrix in a discretized matrix form
\begin{eqnarray}
	\label{eq:lse}
	\mathbb{T} = \mathbb{V} + \mathbb{V}  \mathbb{G}  \mathbb{T},
\end{eqnarray}
where $\mathbb{V}$ and $\mathbb{G}$ are known $n \times n$ matrices and the $n \times n$ matrix $\mathbb{T}$ is to be determined. All matrices in the LSE depend on the energy $E$ which is kept fixed. 
We aim at computing $\mathbb{T}$ for the updated potential (\emph{affine form})
\begin{eqnarray}
  \label{eq:modpot2}
  \mathbb{V} = \mathbb{V}_0 + \sum_{i} c_i \mathbb{V}_i,
\end{eqnarray}
where $\vec c$ are real variable parameters and $\mathbb{V}_i$  known fixed matrices.
We assume that the matrix $\mathbb{T}_0$, which fulfills the LSE  
$\mathbb{T}_0 = \mathbb{V}_0 + \mathbb{V}_0  \mathbb{G} \mathbb{T}_0$, is known. 
Without loss of generality, one can identically rewrite Eq.~\eqref{eq:modpot2} as
\begin{eqnarray}
	\label{eq:modpot}
	\mathbb{V} = \mathbb{V}_0 + \mathbb{X}  \mathbb{C} \mathbb{Z},
\end{eqnarray}
where all parameters $\vec c$ are collected in an $r \times r$ matrix $\mathbb{C}$, while $\mathbb{X}$ and $\mathbb{Z}$ are parameter-independent matrices of size $n \times r$ and $r \times n$, respectively.

In many applications, the matrix $\mathbb{X}  \mathbb{C}  \mathbb{Z}$ has a low rank $r \ll n$. This is, e.g., the case for the two-nucleon (2N) potentials of Refs.~\cite{Epelbaum:2014sza,Reinert:2017usi,Entem:2017gor,Ekstrom:2013kea,Ekstrom:2015rta} derived in chiral EFT, where $c_i$ are given by LECs multiplying 2N contact interactions.   
The well-known Woodbury identity for matrix inversion~\cite{Henderson:1981woodbury} shows that a low-rank update of a matrix leads to a low-rank update of its inverse. Using this identity, we have derived the Woodbury identity for the LSE,
\begin{eqnarray}
    \label{eq:woodlse}
  \mathbb{T} &=& \mathbb{T}_0 + \tilde{\mathbb{X}} \tilde{\mathbb{C}} \tilde{\mathbb{Z}}, \qquad \text{where}
     \\
    \tilde{\mathbb{C}} &\equiv& \big[\mathds{1}_r - \mathbb{C} \mathbb{Z} \mathbb{G}   (\mathbb{T}_0  \mathbb{G} + \mathds{1}_n) \mathbb{X} \big]^{-1}  \mathbb{C}, \nonumber
     \\
     \tilde{\mathbb{X}} &\equiv& (\mathbb{T}_0  \mathbb{G} + \mathds{1}_n)  \mathbb{X}
     \;\;\text{  and  }\;\;
     \tilde{\mathbb{Z}} \equiv \mathbb{Z} (\mathds{1}_n + \mathbb{G}  \mathbb{T}_0). \nonumber
\end{eqnarray}
Equation \eqref{eq:woodlse} is proven in the Supplemental Material \cite{SM} and shows that the update $\tilde{\mathbb{X}}  \tilde{\mathbb{C}} \tilde{\mathbb{Z}}$ of the solution of the LSE has at most rank $r$.
It is important that the parameters $\vec c$ enter only through the $r \times r$ matrix $\tilde{\mathbb{C}}$,
while all other matrices are $\vec c$-independent and can be pre-calculated.
The identity \eqref{eq:woodlse} is valid even for singular $\mathbb{C}$.
For problems where pre-computation of the entire $n \times n$ matrix $\mathbb{T}_0$ is feasible and the rank of update $r \ll n$, Eq.~\eqref{eq:woodlse} provides an exact and very fast method for calculating parameter-dependent updates of the LSE.

To illustrate the power of the LSE Woodbury identity \eqref{eq:woodlse}, we build an exact emulator for 2N phase shifts. Specifically, we consider a 2N S-wave potential $\mathbb{V}^{2N} = \mathbb{V}_0 + c_1 \mathbb{V}_1 + c_2 \mathbb{V}_2+ c_3 \mathbb{V}_3$, which depends on three LECs $c_{1,2,3}$. See Ref.~\cite{SM} for details. The contact interactions of increasing chiral order $\mathbb{V}_1$, $\mathbb{V}_2$ and $\mathbb{V}_3$ have ranks 1, 2 and 1, respectively. The parameter-free part $\mathbb{V}_0$ stems from the one-pion exchange. 
The phase shift $\delta$ at the energy $E = q^2/(2 \mu)$, where $\mu$ is the reduced mass and $q$ the on-shell momentum, is given by $\delta=-\mathrm{arctan}(2 \pi \mu q K)$. Here, $K$ denotes the on-shell value of the real-valued K-matrix, which is defined as the solution of the LSE with the principal value kernel. For the case at hand, the LSE Woodbury formula \eqref{eq:woodlse} becomes a $2\times 2$ matrix equation that can be solved analytically. This allows one to obtain a semi-analytic expression for the on-shell K-matrix $K (c_1, c_2, c_3)$ at a fixed energy in the form
\begin{equation}
    K =  
    K_0 + \frac{a_1 c_2^{2} + a_2 c_2 + a_3 c_1 + a_4 c_1 c_3 + a_5 c_3}
    {a_6 c_2^{2} +a_7 c_2 + a_8 c_1 + a_9 c_1 c_3 + a_{10} c_3 - 1}\,.
    \label{eq:kmatremu}
\end{equation}
The numerical coefficients $a_i$ and $K_0$ depend on the energy $E$, the specific form of $\mathbb{V}^{2N}$ and the mesh points, see Ref.~\cite{SM} for an explicit example. Notice that the absence of terms proportional to $c_1^2$, $c_3^2$, $c_1 c_2$ and $c_3 c_2$ in Eq.~(\ref{eq:kmatremu}) is specific to the considered model.
The solution of a large matrix equation with variable $c_1$, $c_2$, $c_3$ thus reduces to a trivial evaluation of a simple expression \eqref{eq:kmatremu} with pre-computed $a_i$ and $K_0$. The computational cost of the final (online) emulation phase is negligible, yet the result is an exact solution of the original matrix equation. The most computationally expensive step is solving a single $n \times n$ matrix equation to pre-compute the matrix $\mathbb{K}_0$. For the chiral EFT 2N potentials of Ref.~\cite{Reinert:2017usi}, we found a speed-up of $\sim 3000$ times compared to a direct solution of Eq.~(\ref{eq:lse}) using 60 mesh points.

{\it Three-body scattering.---} 
Emulation becomes even more important for computationally demanding three-body scattering. The corresponding Faddeev equation is still of the LSE type, 
but the typical matrix size in a partial-wave basis becomes $\sim 10^5 \times 10^5$. 
Accordingly, the Faddeev equation is usually solved only for the $n \times 1$-dimensional half-shell part of the T-matrix $T^{3N}$ using iterative methods in combination with the Pad\'e resummation technique \cite{Gloeckle:1995jg}.
Such  methods can be used to compute solutions for specific parameter values (snapshots), but are too slow for making fits in large parameter spaces. 

In the following, we consider an emulator for nucleon-deuteron (Nd) scattering at a fixed energy with a parameter-dependent 3N force.\footnote{Our method is, however, equally applicable to emulating 3N scattering with parametric low-rank updates of 2N forces.}
We start with the Faddeev equation in a discretized matrix form (see Ref.~\cite{Gloeckle:1995jg}):
\begin{eqnarray}
\label{eq:Faddeev}
\mathbb{T}^{3N} &=& \big(\tilde{\mathbb{V}} + \mathbb{t} \mathbb{P}\big) +
                    \big(\tilde{\mathbb{V}} + \mathbb{t} \mathbb{P}\big) \mathbb{G} \mathbb{T}^{3N}, \quad \text{where} \\
 \tilde{\mathbb{V}} &\equiv& \left(\mathds{1}_n+\mathbb{t} \mathbb{G}\right) \mathbb{V}^{3N}(\mathds{1}_n+\mathbb{P}), \;\;
 \mathbb{V}^{3N}  \equiv \,  \mathbb{V}_0 + \sum_{i} c_i \mathbb{V}_{c_i}.
 \nonumber 
\end{eqnarray}
Here, the 2N T-matrix $\mathbb{t}$, the free 3N resolvent operator $\mathbb{G}$, the permutation operator $\mathbb{P}$ and the first Faddeev components $\mathbb{V}_0$, $\mathbb{V}_{c_i}$ of the 3N force are known energy-dependent $n \times n$ matrices.  
The $n \times 1$ vector $T^{3N}$ we are interested in is given by $T^{3N} \equiv \mathbb{T}^{3N}\phi$, where
the initial state 
$\phi$ is a known $n \times 1$ vector. 
The equation for obtaining $T^{3N}$ iteratively is found by multiplying Eq.~\eqref{eq:Faddeev} with $\phi$ from right.

We now show how to build an exact and very fast emulator for $T^{3N}(\vec c \, )$ if $\mathbb{V}_i$ are of low rank. 
Since Eq.~\eqref{eq:Faddeev} has the form of the LSE~\eqref{eq:lse} and the matrix $\tilde{\mathbb{V}} + \mathbb{t} \mathbb{P}$ receives a low-rank update, the $n \times n$ parameter-dependent Faddeev equation (\ref{eq:Faddeev}) can be reduced to an $r \times r$ matrix equation with no accuracy loss as discussed in the previous section. While a direct application of the LSE Woodbury identity~\eqref{eq:woodlse} to Eq.~\eqref{eq:Faddeev} requires pre-calculating the $n \times n$ matrix $\mathbb{T}^{3N}_0$, which is not feasible, we can build an exact snapshot-based emulator that only requires the calculation of $r+1$ snapshots.
Indeed, the LSE Woodbury identity~\eqref{eq:woodlse} implies that for any values of $\vec c$,  the vector $T^{3N} = \mathbb{T}^{3N}\phi$ can be expressed as a linear combination of at most $r+1$ $\vec c\,$-independent vectors $T_a^{3N}$,
\begin{eqnarray}
\label{LinComb}
  T^{3N} (\vec c \, ) = T^{3N}_0 + \sum_{a=1}^{r} f_a(\vec c \, )\,  T^{3N}_a.
\end{eqnarray}
To find a reduced basis that spans the whole $(r+1)$-dimensional space of solutions $T^{3N} (\vec c \, )$, it
is sufficient to compute $r+1$ linearly independent snapshots for fixed values of $c_i$ and orthonormalize them. Projecting Eq.~(\ref{eq:Faddeev}) on this reduced basis then yields an $(r+1)$-dimensional matrix equation. 
Remarkably, this basis reduction does not affect the accuracy of the solution, and the emulated result coincides with the exact one.

We now illustrate these ideas by emulating Nd scattering in chiral EFT \cite{Epelbaum:2008ga}, where the leading 3N force has the form  
$\mathbb{V}^{3N} = \mathbb{V}_0 +  c_D \mathbb{V}_{c_D} + c_E \mathbb{V}_{c_E}$. Here, $\mathbb{V}_0$ is the parameter-independent part due to the two-pion exchange, while the short-range 3N force contributions depend on the LECs $c_D$ and $c_E$, $|c_D| \sim |c_E| \sim \mathcal{O} (1)$, see Ref.~\cite{Maris:2020qne} for explicit expressions and Ref.~\cite{SM} for implementation details. 

It is worth emphasizing that the $c_E$-interaction is separable of rank 1, while the $c_D$-part of the 3N force is of non-separable type. 

\begin{figure}[t]
	\centering
	\includegraphics[width=1\linewidth]{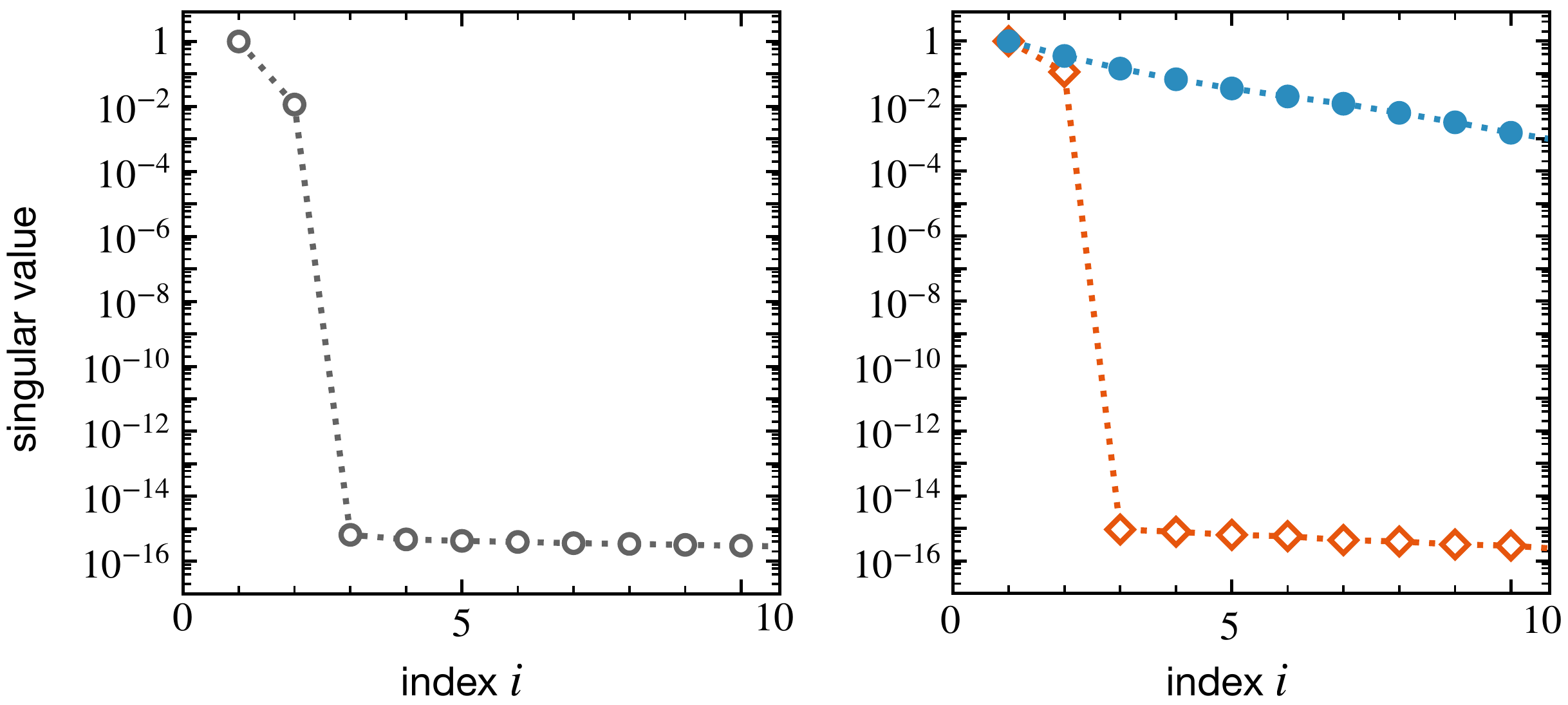}
    \vspace{-0.5cm}
	\caption{
    SVD of solution-vectors $T^{3N}(c_E)$ to the Faddeev equation~\eqref{eq:Faddeev}
    at a fixed energy (left) and of the two-body bound state wave functions using $\mathbb{V}^{2N} (c_1,c_2,c_3)$ (right).
    Orange diamonds (blue circles) correspond to a
    setup where the binding energy is (is not) constrained as explained in the text.
    }
	\label{SVD_snapshotmatrix}
     \vspace{-0.2cm}
\end{figure}

We first build a single-parameter emulator for $T^{3N} (c_E)$ with a constant $c_D$ term absorbed into $\mathbb{V}_0$ \cite{SM}.
Throughout this section, we fix the energy to $E_{\rm lab} = 70$~MeV. 
For the considered rank-1 update of $\mathbb{V}^{3N}$, the solution $T^{3N} (c_E)$ must, according to Eq.~(\ref{LinComb}), be expressible as a linear combination of just two vectors $T^{3N}_0$ and $T^{3N}_1$. This can be verified numerically by computing snapshots for multiple values of $c_E$ and making singular value decomposition (SVD) of the matrix composed of these snapshots, see the left panel of Fig.~\ref{SVD_snapshotmatrix}. 
It is thus indeed sufficient to compute any two linearly independent snapshots using different values of $c_E$ to build a reduced basis. By projecting the Faddeev equation for $T^{3N}$ onto the orthonormalized reduced basis vectors, the original $10^5 \times 10^5$ matrix equation (\ref{eq:Faddeev}) reduces to a trivial $2 \times 2$ matrix equation without accuracy loss.\footnote{By solving this matrix equation one can, if desired, obtain semi-analytic expressions for 3N scattering observables in terms of $c_E$ in a close analogy to Eq.~(\ref{eq:kmatremu}).} 
\begin{figure}[t]
	\centering
	\includegraphics[width=\linewidth]{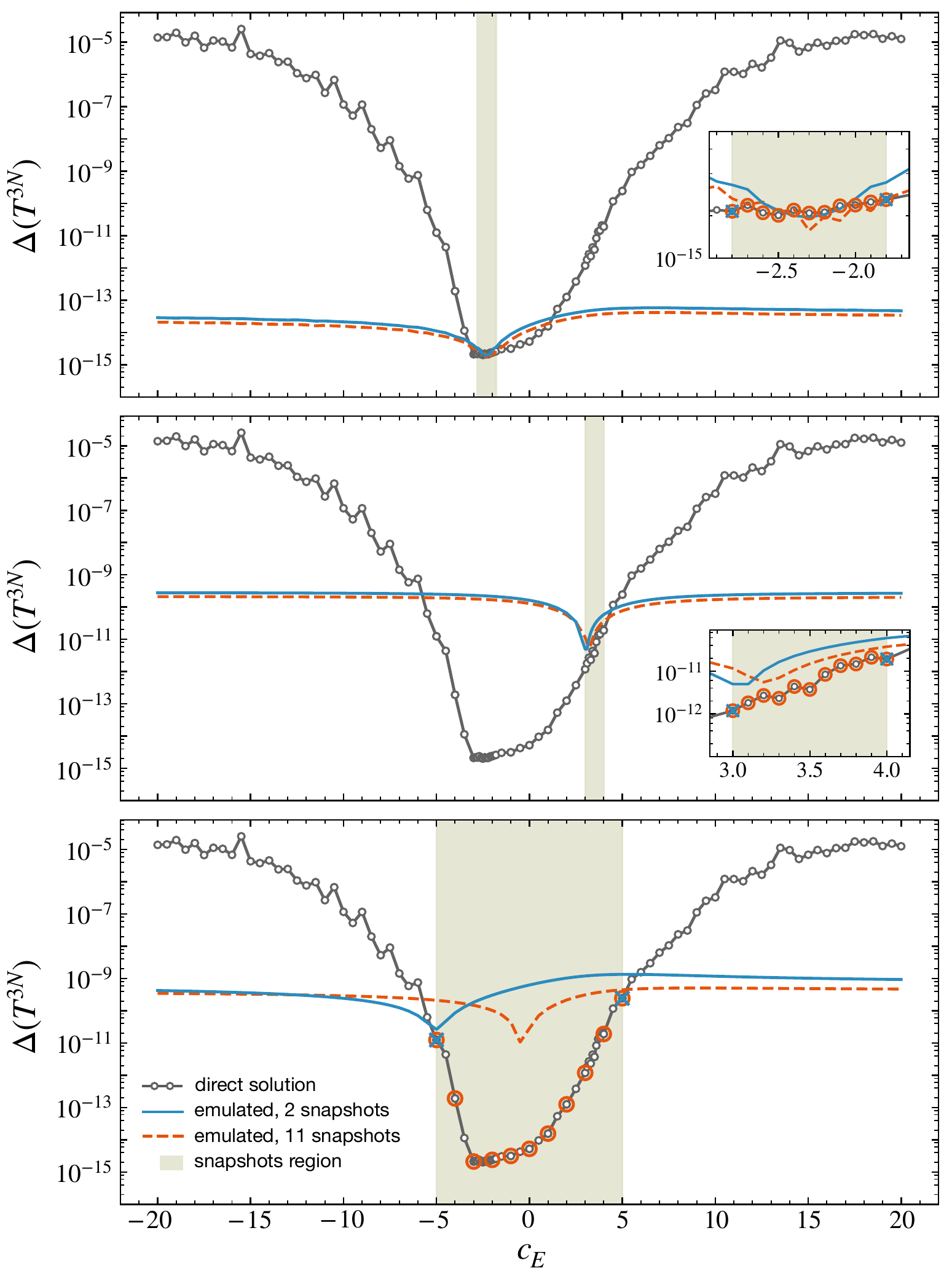}
        \vspace{-0.5cm}
	\caption{
Numerical accuracy of the direct and emulated solutions $T^{3N} (c_E)$ in the $J^\pi=\frac{1}{2}^+$ channel. Gray open circles correspond to direct solutions of the Faddeev equation obtained via Pad\'e resummation of 30 iterations. Blue solid (red dashed) lines show the accuracy of the emulated result using 2 (11) snapshots, whose locations are indicated by blue crosses (red circles). Light-shaded regions mark the range of $c_E$-values used to generate snapshots. 
}
	\label{prec_differentcE_emulators}
        \vspace{-0.2cm}
\end{figure}

It is important to address the impact of the number and accuracy of snapshots on the emulator accuracy. In Fig.~\ref{prec_differentcE_emulators}, we show the numerical error $\Delta (T^{3N})$, defined via 
\begin{equation}
    \Delta(T^{3N}) \equiv \frac{\lVert ( \tilde{\mathbb{V}} + \mathbb{t} \mathbb{P})\phi+
                    ( \tilde{\mathbb{V}} + \mathbb{t} \mathbb{P}) \mathbb{G} T^{3N}-T^{3N}\rVert}{\lVert T^{3N} \rVert},
                    \label{eq:numerrdef}
\end{equation}
where $\lVert X \rVert$ denotes a two-norm of $X$, for the exact and emulated results in a wide range of $c_E$-values for different snapshot choices. 
\begin{figure*}[t!]
	\centering
	\includegraphics[width=.92\linewidth]{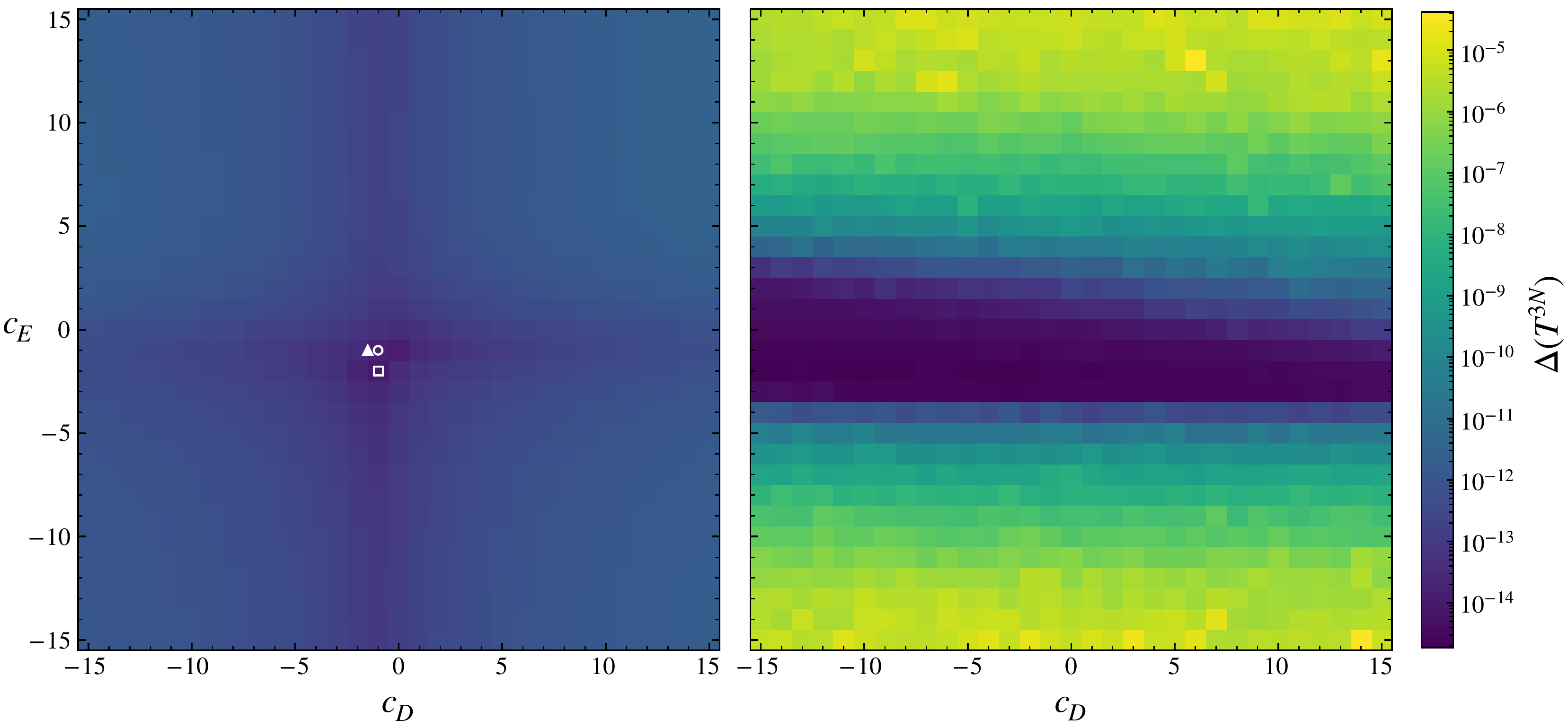}
        \vspace{-0.2cm}
	\caption{
    Numerical accuracy of the emulated (left) and direct (right) solutions $T^{3N} (c_D, c_E)$ in the $J^\pi=\frac{1}{2}^+$ channel.
    Direct solutions are obtained by performing Pad\'e resummation of 40 iterations of the Faddeev equation.
 White open circle, open square and filled triangle mark the parameter values used to generate the main snapshot, the $\mathbb{V}_{c_E}$-snapshot and the 28 snapshots for each of the rank-1 component of  $\mathbb{V}_{c_D}$, respectively. 
 }
	\label{prec_heatmap}
        \vspace{-0.2cm}
\end{figure*}
In an ideal situation, snapshots  may be expected to be accurate to machine precision. However, as already mentioned, the 
numerical method for solving the Faddeev equation uses Pad\'e resummation of the Neumann series. The accuracy of snapshots calculated this way deteriorates when $|c_E|$ becomes large, see gray open circles in Fig.~\ref{prec_differentcE_emulators}. For the emulated results, we observe that two accurate snapshots, which can be chosen close to each other, are fully sufficient for building an accurate emulator (for the considered rank-1 update), and its accuracy 
is governed by the accuracy of the least precise snapshot.  
Increasing the number of snapshots does not improve the emulation accuracy and may even  degrade the emulator quality if the additional snapshots are less accurate.
We emphasize that our emulator does not generate any extrapolation artifacts and yields precise solutions of the Faddeev equation even for values of $c_E$ far away from the snapshot region. Even more remarkable is that the accuracy of the emulated result by far exceeds the accuracy of the direct solution for large values of $|c_E|$. 
This shows that such an emulator can be used as a resummation tool to obtain solutions for parameter values where conventional methods do not converge.

To illustrate the treatment of updates with rank higher than 1, we now emulate the solution $T^{3N} (c_D, c_E)$ for the full 3N force. While the $c_D$-interaction is non-separable, the matrix $\mathbb{V}_{c_D}$ has rank 28 for the employed discretization setup \cite{SM}. Since for rank-1 updates, the accuracy of the emulator does not depend on the choice of snapshots (provided they are numerically accurate), it is advantageous to represent $\mathbb{V}_{c_D}$ as a sum of 28 rank-1 potentials using SVD: $\mathbb{V}_{c_D} = \sum_a \sigma_a u_a v^{*}_a \equiv  \sum_a \mathbb{V}_{c_D,\, a}^{\text{rank-1}}$. We then first compute the main snapshot $T_0^{3N}$ using, e.g., $\mathbb{V}^{3N} = \mathbb{V}_0$, and set successively $\mathbb{V}^{3N} = \mathbb{V}_0 + c_D \mathbb{V}_{c_D,\, a}^{\text{rank-1}}$ with $a=1, \ldots, 28$ and any $c_D \neq 0$ to generate 28 snapshots $T^{3N}_{1 \dots 28}$. Finally, we compute the snapshot $T_{29}^{3N}$ using $\mathbb{V}^{3N} = \mathbb{V}_0 + c_E \mathbb{V}_{c_E}$ and a nonzero value of $c_E$. After building an orthonormal basis from 30 vectors $T^{3N}_{0 \dots 29}$ and projecting Eq.~(\ref{eq:Faddeev}) on this basis, the dimensionality of the $10^5 \times 10^5$ Faddeev equation reduces to  $30 \times 30$ without accuracy loss. 

Unlike typical snapshot-based emulators, which use sophisticated algorithms for choosing snapshots in order to optimize accuracy  \cite{Sarkar:2021fpz,Gnech:2025lbg,Maldonado:2025ftg}, SVD of the Hamiltonian update into rank-1 terms allows us to prepare snapshots for each rank-1 update independently of each other and of the parameter values.
This way our method provides an accurate, unambiguous, and scalable algorithm for building a complete reduced bases.

In Fig.~\ref{prec_heatmap}, we show the snapshot locations of our emulator and compare numerical errors for the emulated (left panel) and directly computed (right panel) solutions $T^{3N}$ for a wide range of $c_D$-, $c_E$-values. Similarly to the $T^{3N} (c_E)$ case, the emulated results do not show any error increase outside the snapshot region (as expected for an exact emulator) and are much more accurate than the iterative solution for large values of $|c_D|$, $|c_E|$. According to our tests, the achieved speed up of the  $T^{3N} (c_D, c_E)$-emulator amounts to $\sim 10^6$. 

{\it Bound states.---}
The methodology described in the previous sections can be used to efficiently constrain nuclear interaction models from scattering data. However, there is also growing interest in tuning nuclear forces to properties of nuclei like, e.g., the ground- and excited-state energies, charge radii, and decay rates \cite{Ekstrom:2013kea,Ekstrom:2015rta,Wesolowski:2021cni,Sun:2024iht}. This requires a computationally efficient emulation of bound-state wave functions for low-rank updates of the Hamiltonian. We now show how this can be achieved using the RBM introduced above. 

We first emphasize that the LSE Woodbury identity \eqref{eq:woodlse} can, in principle, be directly applied to efficiently constrain $\vec c \, $ such that the system develops a bound state at any given energy $\bar E$. The requirement of the T-matrix to have a pole at $E= \bar E$ yields the  condition $\det(\mathbb{D}) = 0$, with the $r \times r$ matrix $\mathbb{D} \equiv  \mathds{1}_r - \mathbb{C} \mathbb{Z} \mathbb{G}   (\mathbb{T}_0  \mathbb{G} + \mathds{1}_n) \mathbb{X}$. Moreover, for $\det(\mathbb{D}) = 0$, $\text{adj} (\mathbb{D})$ has rank 1, meaning that its column space has dimension 1,
and $\Psi^{\bar E} (\vec c\,) \equiv \text{colsp}\big[\text{adj}(\mathbb{D})\big] \mathbb{C} \tilde{\mathbb{Z}}$ yields the (non-normalized) bound-state wave function. This can be used to emulate bound-state wave functions for values of $\vec c\, $ that preserve the binding energy. 

In practical applications, it is more convenient to use an approach that does not require calculating the T-matrix. Consider the matrix eigenvalue problem 
\begin{eqnarray}
  \mathbb{H} \Psi^{E} &=& E \Psi^{E} ,\quad \text{with}
  \;\; \mathbb{H} = \mathbb{H}_0 + \sum_{i} c_i \mathbb{V}_i\,,
  \label{eq:updatedse}
\end{eqnarray}
where $\mathbb{H}_0$ and $\mathbb{V}_i$ are known $n \times n$ matrices and $\Psi^E$ the corresponding eigenvector. Let $E = \bar E$ denote a non-degenerate discrete eigenvalue of the Hamiltonian $\mathbb{H}$ corresponding to some parameter values $c_i = \bar c_i$. 
Our goal is to emulate $\Psi^{\bar E} (\vec c \,)$ for a parameter-dependent update of the Hamiltonian, subject to the constraint $E (\vec c\, ) = \bar E$. This can be achieved using the result proven in the Supplemental Material \cite{SM}, which shows that $\Psi^{\bar E} (\vec c \,)$ can be written as a linear combination
\begin{eqnarray}
  \Psi^{\bar E} (\vec c\, ) = \sum_{a=1}^{r} f_a(\vec c\,) \, \Psi^{\bar E}_a,
\end{eqnarray}
where $\Psi^{\bar E}_a$ are $\vec c$-independent reduced-basis vectors and $r$
does not exceed the combined rank of $\mathbb{V}_i$. We emphasize that this dimensionality reduction of the Schr\"odinger equation is only valid for updates that preserve the binding energy. 
This is illustrated in the right panel of Fig.~\ref{SVD_snapshotmatrix}, which shows SVD of 2N bound-state wave functions for the interaction model $\mathbb{V}^{2N}$, see Ref.~\cite{SM} for details, where the parameters $c_{1,2,3}$ are either chosen completely randomly (blue circles) or constrained to reproduce the binding energy set to $\bar E=-2.22$~MeV (orange diamonds). 

We now present the algorithm for building an emulator for the $A$-body wave function $\Psi^{\bar E} (\vec c \,)$:
\vspace{-0.2cm}
\begin{enumerate}[itemsep=0.2ex, parsep=0.4ex, leftmargin=0.6cm]
\item
Perform (randomized) SVD of $\mathbb{V}_i$'s in Eq.~\eqref{eq:updatedse} to write them as a sum of rank-1 matrices, $\mathbb{V}_i = \sum_{a} \mathbb{V}_{i, a}^{\text{rank-1}}$.
\item
For each $\mathbb{V}_{i,a}^{\text{rank-1}}$, tune the value of $c_i$ to ensure that
  $(\mathbb{H}_0 + c_i \mathbb{V}_{i,a}^{\text{rank-1}}) \Psi_{i,a}^{\bar E} = \bar E \,\Psi_{i,a}^{\bar E}$ and
  save all $\Psi_{i,a}^{\bar E}$. For rank-1 updates, such $c_i$ always exist and are unique.
\item 
Build an orthonormal basis from all eigenvectors $\Psi_{i,a}^{\bar E}$ obtained in step 2.
\item 
Project the full Hamiltonian $\mathbb{H}$ from Eq.~\eqref{eq:updatedse} onto this reduced basis to obtain $\mathbb{H}_\text{rb}$.
\item
In the reduced basis, find a constraint on the parameters $c_i$ to keep the eigenvalue of $\mathbb{H}_\text{rb}$ equal to $\bar E$.
\item 
In the reduced basis, for any $\vec c$ that fulfill the constraint from step 5, compute the eigenvector of 
$\mathbb{H}_\text{rb}$.
\item
Convert it to the original basis to obtain $\Psi^{\bar E} (\vec c \,)$.
\end{enumerate}
\vspace{-0.2cm}

The resulting emulator allows one to calculate any desired bound-state observable for values of $\vec c \, $ consistent with the fixed binding energy $\bar E$ without computational effort. As a proof-of-principle, we describe in Ref.~\cite{SM} an emulator for the ``deuteron'' wave function $\Psi^{\bar E} (c_1, c_2, c_3)$ using the already introduced potential model $\mathbb{V}^{2N}$ and provide a semi-analytic expression for the mean-square matter radius $r_m^2 (c_1, c_2, c_3)$.

{\it Summary.---}
We have proven that for a parametric low-rank update of the Hamiltonian, the $A$-body scattering and bound-state problems with a fixed energy {\it exactly} reduce to a low-dimensional matrix equation whose solution is trivial from the computational point of view. This explains the empirically observed good performance of various snapshot-based emulators \cite{Witala:2021xqm,Wesolowski:2021cni,Bonilla:2022rph,Garcia:2023slj,Sun:2024iht,Armstrong:2025tza,Gnech:2025lbg}. We describe algorithms for building {\it exact} snapshot-based emulators for scattering and bound-state wave functions and provide explicit examples for $A=2,\, 3$ systems. Using SVD of the Hamiltonian update, we have shown that choosing optimal snapshot points in the parameter space becomes unambiguous with no need for advanced snapshot finding algorithms \cite{Sarkar:2021fpz,Gnech:2025lbg,Maldonado:2025ftg}. Owing to the exactness of our reduction, we are able to emulate observables for parameter values far away from the snapshot region without accuracy loss, a feature that distinguishes our approach from the existing emulators. For very low-rank updates, our method can even be used to derive simple yet numerically exact semi-analytical expressions for observables in terms of the parameters of the Hamiltonian.
We have also demonstrated the ability of our emulators to serve as a resummation tool by providing accurate results in those regions of the parameter space, which cannot be accessed using conventional solution techniques. 

The methods described in this paper are not limited to particular interactions and can be applied to a wide range of problems. They will be indispensable for establishing accurate and precise 3N forces in chiral EFT. More generally, our results open new perspectives in computational nuclear physics and may have applications in quantum chemistry, atomic physics and other areas dealing with the quantum many-body problem.

\begin{acknowledgments}
We are grateful to Henryk Wita\l{}a for a long-standing collaboration on research focused on 3N scattering and for sharing with us his experience and codes. We are also grateful to Josep Sol\`{a} Cava for collaborating during the early stage of this project and to all members of the LENPIC Collaboration for sharing their insights into the considered topics. This work has been supported by the European Research Council (ERC) under the European Union’s Horizon 2020 research and innovation programme (grant agreement No.~885150), by the MKW NRW under the funding code NW21-024-A, by JST ERATO (Grant No. JPMJER2304), by JSPS KAKENHI (Grant No. JP20H05636) and by BMBF through the ErUM-Data project DEMOS.
\end{acknowledgments}


\appendix
\renewcommand{\theequation}{\thesection.\arabic{equation}}
 
\bibliographystyle{unsrturl}
\bibliography{bibliography.bib}

\beginsupplement
\onecolumngrid

\newpage

\section{Supplemental Material}

This Supplemental Material is organized as follows. We start with providing in Sec.~I a detailed derivation of the Lippmann-Schwinger Woodbury identity (\ref{eq:woodlse}), followed by the proof of dimensional reduction of the Schr\"odinger equation for bound states in Sec.~II. Next, in Sec.~III, we specify  the two-body interaction model and describe in detail the construction of the two-body scattering and bound-state emulators introduced in the main text. The corresponding Jupyter Notebooks are available as ancillary files. 
Finally, in Sec.~IV, we specify the interactions used to emulate nucleon-deuteron scattering and provide details on the numerical solution of the Faddeev equation. 

\subsection{I.~Proof of the Woodbury identity (\ref{eq:woodlse}) for the LSE}

In this section, we prove the identity Eq.~\eqref{eq:woodlse}, which demonstrates that a low-rank update of the potential in the LSE leads to a low-rank update of its solution.
First, we consider the LSE in matrix form before updating the potential:
\begin{eqnarray}
	\label{eq:lse0}
\mathbb{T}_0 = \mathbb{V}_0 + \mathbb{V}_0  \mathbb{G} \mathbb{T}_0.
\end{eqnarray}
We collect all terms that involve the parameter-independent T-matrix $\mathbb{T}_0$ on the left-hand side
and obtain a system of linear equations
\begin{eqnarray}
	\label{eq:t0system}
	(\mathds{1}_n -  \mathbb{V}_0 \mathbb{G})  \mathbb{T}_0 = \mathbb{V}_0.
\end{eqnarray}
The solution $\mathbb{T}_0$ can be found 
by multiplying both sides of Eq.~\eqref{eq:t0system} by an inverse matrix:
\begin{eqnarray}
	\mathbb{T}_0 = {(\mathds{1}_n -  \mathbb{V}_0 \mathbb{G})}^{-1} \mathbb{V}_0.
\end{eqnarray}
For brevity, we introduce an auxiliary $n \times n$ matrix $\mathbb{A}$, defined as
\begin{eqnarray}
	\mathbb{A} \equiv (\mathds{1}_n -  \mathbb{V}_0 \mathbb{G}).
\end{eqnarray}
The solution of Eq.~\eqref{eq:lse0} can then be written as
\begin{eqnarray}
	\label{eq:simplifLSE0}
	\mathbb{T}_0 = \mathbb{A}^{-1} \mathbb{V}_0.
\end{eqnarray}
Next, we note that the inverse matrix $\mathbb{A}^{-1}$ satisfies the identity
\begin{eqnarray}
	\label{eq:invaproperty}
	\mathbb{A}^{-1} = \mathbb{T}_0 \mathbb{G} + \mathds{1}_n.
\end{eqnarray}
This can be shown by multiplying both sides of the above equation 
with the matrix $\mathbb{A}$ from the left and verifying that the result
is equivalent to the Lippmann-Schwinger equation~\eqref{eq:lse0}.

Next, we consider the LSE \eqref{eq:lse} for the updated potential \eqref{eq:modpot}.
Its solution can be written as:
\begin{eqnarray}
	\mathbb{T} &=& {(\mathds{1}_n -  \mathbb{V}  \mathbb{G})}^{-1}   \mathbb{V}
	\nonumber
	\\
	&=& {(\mathds{1}_n -  \mathbb{V}_0  \mathbb{G} - \mathbb{X}  \mathbb{C}  \mathbb{Z}  \mathbb{G})}^{-1}   (\mathbb{V}_0 + \mathbb{X}  \mathbb{C}  \mathbb{Z})
	\nonumber
	\\
	&=&
	{(\mathbb{A} - \mathbb{X}  \mathbb{C}  \mathbb{Z} \mathbb{G})}^{-1}  (\mathbb{V}_0 + \mathbb{X} \mathbb{C}  \mathbb{Z}).
	\label{eq:lsesol1}
\end{eqnarray}
Since the matrix $(- \mathbb{X})  \mathbb{C}  (\mathbb{Z}  \mathbb{G})$ 
may be considered as a low-rank update of the matrix $\mathbb{A}$,
we can apply the original Woodbury identity for the inverse matrix in Eq.~\eqref{eq:lsesol1}.
In general, the Woodbury identity for update of matrix inverse reads
\begin{eqnarray}
	\label{eq:woodorig}
	(\mathbb{A} + \mathbb{U}  \mathbb{C}  \mathbb{W})^{-1}
	= \mathbb{A}^{-1}
	- \mathbb{A}^{-1} 
	 \mathbb{U} 
	 (\mathds{1}_r + \mathbb{C}  \mathbb{W}  \mathbb{A}^{-1}  \mathbb{U})^{-1}
	 \mathbb{C}
	 \mathbb{W}
	 \mathbb{A}^{-1},
\end{eqnarray}
where
$\mathbb{A}$, $\mathbb{U}$, $\mathbb{C}$, and $\mathbb{W}$ are matrices of the size $n \times n$, $n \times r$, $r \times r$, and $r \times n$, respectively. 
In our case, $\mathbb{U} = (- \mathbb{X})$ and $\mathbb{W} = \mathbb{Z} \mathbb{G}$.
After applying the Woodbury identity, Eq.~\eqref{eq:lsesol1} transforms into
\begin{eqnarray}
	\mathbb{T} =
	\left[
		\mathbb{A}^{-1}
		+ \mathbb{A}^{-1} 
		 \mathbb{X} 
		 (\mathds{1}_r - \mathbb{C}  \mathbb{Z}  \mathbb{G}  \mathbb{A}^{-1}  \mathbb{X})^{-1}
		 \mathbb{C}
		 \mathbb{Z}
		 \mathbb{G}
		 \mathbb{A}^{-1}
	\right]
	 (\mathbb{V}_0 + \mathbb{X}  \mathbb{C}  \mathbb{Z}).
	\label{eq:lsesol2}
\end{eqnarray}
Next, we expand the brackets and add $(- \mathds{1}_r + \mathds{1}_r)$  to the last term:
\begin{eqnarray}
	\mathbb{T} &=&
	\mathbb{A}^{-1}  \mathbb{V}_0
	\nonumber
	\\
	&&+
		\mathbb{A}^{-1} 
		 \mathbb{X} 
		 (\mathds{1}_r - \mathbb{C}  \mathbb{Z}  \mathbb{G}  \mathbb{A}^{-1}  \mathbb{X})^{-1}
		 \mathbb{C}
		 \mathbb{Z}
		 \mathbb{G}
		 \mathbb{A}^{-1}
		 \mathbb{V}_0
	\nonumber
	\\
	&&+ \mathbb{A}^{-1} \mathbb{X}  \mathbb{C} \mathbb{Z}
	\nonumber
	\\
	&&+
		\mathbb{A}^{-1} 
		 \mathbb{X} 
		 (\mathds{1}_r - \mathbb{C} \mathbb{Z} \mathbb{G}  \mathbb{A}^{-1}  \mathbb{X})^{-1}
		 (\mathbb{C}
		 \mathbb{Z}
		 \mathbb{G}
		 \mathbb{A}^{-1}
		 \mathbb{X} - \mathds{1}_r + \mathds{1}_r)  \mathbb{C}  \mathbb{Z}.
\end{eqnarray}
The resulting expression can be further simplified to
\begin{eqnarray}
	\mathbb{T} =
	\mathbb{A}^{-1}  \mathbb{V}_0
	+
		\mathbb{A}^{-1} 
		 \mathbb{X} 
		 (\mathds{1}_r - \mathbb{C}  \mathbb{Z}  \mathbb{G}  \mathbb{A}^{-1}  \mathbb{X})^{-1}
		 \mathbb{C}
		 \mathbb{Z}
		 ( \mathds{1}_n + \mathbb{G}
		 \mathbb{A}^{-1}
		 \mathbb{V}_0).
\end{eqnarray}
Finally, we express all inverse matrices $\mathbb{A}^{-1}$ in terms of $\mathbb{T}_0$ using
Eqs.~\eqref{eq:simplifLSE0} and \eqref{eq:invaproperty}
and obtain the Woodbury-type identity Eq.~\eqref{eq:woodlse}.

\subsection{II.~Dimensional reduction of the Schrödinger equation for bound states}
\label{subsec:dimreductSE}
 We now consider the eigenvalue problem for the Hamilton operator in matrix form 
 \begin{eqnarray}
 \label{EqStarting}
  \mathbb{H} \Psi^{E} &=& E \Psi^{E} ,\quad \text{with}
  \;\; \mathbb{H} = \mathbb{H}_0 + \mathbb{X}\mathbb{C}\mathbb{Z},
\end{eqnarray}
where $\mathbb{H}_0$, $\mathbb{X}$, and $\mathbb{Z}$ are matrices of the size $n \times n$, $n \times r$, and $r \times n$, respectively.
The $r \times r$ matrix $\mathbb{C}$ and the $n$-dimensional vector $\Psi^E$ depend on the variable parameters $\vec c$. We emphasize again that rewriting Eq.~(\ref{eq:updatedse}) in the form of Eq.~(\ref{EqStarting}) is always possible (without loss of generality).
We consider an arbitrary variation of the parameters $\vec c \,$, constrained by the condition that the discrete eigenenergy $E$ is fixed to 
some non-degenerate value $\bar E$.
That is, we impose the condition $E(\vec c\, ) = \bar E$ and require that this equation possesses solutions and that it is not trivially fulfilled for all possible values of $\vec c$. 

We first introduce a parameter-independent matrix $\mathbb{A} \equiv \mathbb{H}_0 - \bar E \mathds{1}_n$
to cast the eigenvalue equation into the form
\begin{align}
\label{eq:AXCZ}
\mathbb{A} \Psi^{\bar E}(\vec c \, ) = - \mathbb{X} \mathbb{C} \mathbb{Z} \Psi^{\bar E}( \vec c \, ). 
\end{align}
If the matrix $\mathbb{A}$ is invertible, we can apply its inverse $\mathbb{A}^{-1}$ from the left to obtain
\begin{equation}
    \Psi^{\bar E}(\vec c \, ) \, =\,  - \mathbb{A}^{-1}\mathbb{X} \mathbb{C} \mathbb{Z} \Psi^{\bar E}( \vec c \, ) \quad \Rightarrow \quad
    \Psi^{\bar E}(\vec c \, ) \, \in \, \text{Im}(\mathbb{A}^{-1}\mathbb{X}).
\end{equation}
Since the rank of the $n \times r$ matrix $\mathbb{A}^{-1}\mathbb{X}$ is less than or equal to $r$, we conclude that the subspace spanned by $\Psi^{\bar E}(\vec{c}\,)$ has no more than $r$ dimensions.

The matrix $\mathbb{A}$ can also be non-invertible. However, the combined columns of the matrices $\mathbb{A}$ and $\mathbb{X}$ still span the entire space. Accordingly, it is always possible to find an $r \times n $ matrix $\mathbb{F}$, for which the matrix $\mathbb{A}+\mathbb{X}\mathbb{F}$ is invertible. 
Then, one can add $\mathbb{X}\mathbb{F}\Psi^{\bar E}(\vec{c}\,)$ to both sides of Eq.~(\ref{eq:AXCZ}) to obtain
\begin{equation}
\big(\mathbb{A}+\mathbb{X}\mathbb{F}) \Psi^{\bar E}(\vec c \, ) \, =\,  - \mathbb{X}\big( \mathbb{C} \mathbb{Z} +\mathbb{F}) \Psi^{\bar E}( \vec c \, )\quad \Rightarrow \quad
    \Psi^{\bar E}(\vec c \, ) \, \in \,  \text{Im}\big(\big[\mathbb{A}+\mathbb{X}\mathbb{F}\big]^{-1}\mathbb{X}\big).
\end{equation}
Once again, one concludes that the subspace, in which $\Psi^{\bar E}(\vec{c}\,)$ varies due to the constrained variation of the parameters $\vec{c}\, $, has at most $r$ dimensions. This dimensional reduction is independent of the size $n$ of the Hamiltonian matrix.

\subsection{III.~Emulation of two-body observables}

\subsubsection{A.~Two-body interaction}
In this section, we define the interaction model we use to emulate two-body scattering and bound-state observables. The model we employ is a simplified version of the two-nucleon potential derived in chiral EFT and consists of a parameter-free long-range and parameter-dependent short-range interaction.
The long-range part is taken as an S-wave projection of the one-pion exchange two-nucleon potential. In the plane-wave basis, it has the form 
\begin{equation}
V_{\rm long-range} (\vec p\, ', \, \vec p\, ) = - \frac{1}{(2 \pi)^3} {\left( \frac{M g_A}{2 F_\pi} \right)}^2\frac{1}{(\vec p \, ' - \vec p\, )^2 + M^2}, 
\end{equation}
where $\vec p$ and $\vec p \,'$ denote the incoming and outgoing center-of-mass momenta of the interacting particles, $M$ is the pion mass, while $g_A$ and $F_\pi$ are the nucleon axial-vector and pion decay constants, respectively. The short-range part is given by separable contact interactions $V_1$, $V_2$, $V_3$ with adjustable dimensionless coefficients $c_1$, $c_2$, $c_3$, collectively denoted as $\vec{c}$. The considered interaction model has an affine dependence on the adjustable parameters and can be written as
\begin{equation}
V(\vec{c}\, ) = V_0 + c_1 V_1 + c_2 V_2 + c_3 V_3 
\end{equation}
with $V_0 = V_{\rm long-range}$. After projection on the S-wave, it acquires the form  
\begin{equation}
V (p', \, p, \, \vec{c}\, ) = -\frac{1}{8 \pi^2 p' p} {\left( \frac{M g_A}{2 F_\pi} \right)}^2\, \ln \bigg( \frac{(p'+p)^2 + M^2}{(p'-p)^2 + M^2} \bigg) \; + \; \Big( c_1 u_1 + c_2 u_2 (p'{}^2 + p^2) + c_3 u_3 p'{}^2p^2 \Big) e^{-(p'{}^2 + p^2)/\Lambda^2}\,.
\label{eq:potentialswave}
\end{equation}
Here, the cutoff parameter $\Lambda$ controls the range of the short-range interaction. We employ the following numerical values for various (fixed) parameters: $M=138\,\text{MeV}$, $g_A=1.29$, $F_\pi=92.4\,\text{MeV}$, and $\Lambda=500\,\text{MeV}$. The coefficients $u_1 = 10^4 {(2\pi)}^{-3} \text{GeV}^{-2}$,
$u_2 = 10^4 {(2\pi)}^{-3} \text{GeV}^{-4}$, and 
$u_3 = 10^4 {(2\pi)}^{-3} \text{GeV}^{-6}$ in Eq.~(\ref{eq:potentialswave}) are introduced to make the parameters $c_i$ dimensionless with the expected natural size of  $|c_i| \sim \mathcal{O} (1)$. 

To obtain a numerical solution of the Schr{\"o}dinger or Lippmann-Schwinger equation, the potential in Eq.~\eqref{eq:potentialswave} needs to be discretized 
on a mesh of initial and final momenta. 
For simplicity, the meshes for both momenta are taken to be the same.
The elements of the discretized potential matrix $\mathbb{V}$ are defined as 
$[\mathbb{V}]_{ij} = V(p_i, p_j)$,
where $p_i$ are discrete mesh points.
Using this discretization scheme one can rewrite the potential Eq.~\eqref{eq:potentialswave} as a sum of four matrices
\begin{equation}
\mathbb{V}(\vec{c}\, ) = \mathbb{V}_0 + c_1 \mathbb{V}_1 + c_2 \mathbb{V}_2 + c_3 \mathbb{V}_3.
\label{eq:toypotentialmatrix}
\end{equation}
Since the functions $V_{1,2,3}$ correspond to separable interactions, the resulting matrices 
$\mathbb{V}_{1,2,3}$ have very low ranks. 
Specifically, $\mathbb{V}_{1}$ and   $\mathbb{V}_{3}$ have each rank $1$,
while  $\mathbb{V}_{2}$ has rank $2$. The combined rank of all parameter-dependent parts of the potential can, therefore, not exceed $4$. By identically rewriting the parameter-dependent potential as a product of three matrices, it is easy to see that the combined rank actually equals $2$:
\begin{equation}
    \Big( c_1 u_1 + c_2 u_2 (p'{}^2 + p^2) + c_3 u_3 p'{}^2p^2 \Big) e^{-(p'{}^2 + p^2)/\Lambda^2} 
    \equiv
    \left( \begin{array}{cc} e^{-p'{}^2/\Lambda^2}  & p'^2 e^{-p'{}^2/\Lambda^2} \end{array} \right)
    \left( \begin{array}{cc} c_1 u_1 & c_2 u_2\\ c_2 u_2 & c_3 u_3\end{array} \right)
    \left( \begin{array}{cc} e^{-p{}^2/\Lambda^2}  \\ p^2 e^{-p{}^2/\Lambda^2} \end{array} \right),
\end{equation}
where the whole $p'$- ($p$-) dependence is collected in the first (last) matrix, while the parameters $\vec{c}\, $ only enter the $2 \times 2$ matrix.
This identity allows us to rewrite the discretized parameter-dependent part of the potential $c_1 \mathbb{V}_1 + c_2 \mathbb{V}_2 + c_3 \mathbb{V}_3$ in the form of a matrix product $\mathbb{X}\mathbb{C}\mathbb{Z}$, where the elements of the $n \times 2$ matrix $\mathbb{X}$ are given by $[\mathbb{X}]_{i1}=\mathrm{exp}(-p_i^2/\Lambda^2)$ and $[\mathbb{X}]_{i2}=\mathrm{exp}(-p_i^2/\Lambda^2)p_i^2$, while  the $ 2 \times n$ matrix $\mathbb{Z}$ is given by $\mathbb{Z}=\mathbb{X}^T$. The parameter-dependent $2 \times 2$ matrix $\mathbb{C}$ has the form
\begin{equation}
\mathbb{C} = \Big( \begin{array}{cc} c_1 u_1 & c_2 u_2 \\ c_2 u_2 & c_3 u_3\end{array} \Big).
\end{equation}
Since the matrix $\mathbb{C}$ has dimensions $2 \times 2$, the combined rank of the potential update $\mathbb{X}\mathbb{C}\mathbb{Z}$ equals 2.
The full potential matrix acquires the form of Eq.~\eqref{eq:modpot}.

\subsubsection{B.~Scattering emulator}

We now specify the two-body scattering problem, provide details on the discretization 
and demonstrate that our exact emulator yields the same result as obtained by a direct numerical solution of the matrix equation, but in computationally much more efficient way. 

We are interested in calculating the K-matrix, which is a solution of the Lippmann-Schwinger integral equation
\begin{equation}
K (p', \, p, \, E_{\text{cms}}) = V (p', \, p) +2 \mu\,\dashint_0^\infty \mathrm{d}k \,k^2\frac{V (p', \, k)K (k, \, p, \, E_{\text{cms}})}{q^2-k^2},
\label{eq:lsefork}
\end{equation}
where $\mu=470\,\text{MeV}$ is a reduced mass of the two scattered particles (e.g., proton and neutron), $q$ is the center-of-mass momentum with $q^2/(2\mu) = E_{\text{cms}}$ and $\dashint$ denotes the Cauchy principal-value integral.
The interaction potential $V (p', \, p)$ is defined in Eq.~\eqref{eq:potentialswave}.
Our goal is to quickly compute the on-shell K-matrix $K (q, \, q, \, q^2/2\mu)$ as a function of the parameters $\vec{c}$ for fixed energy $E_{\text{cms}}$. In this example,  we take $E_{\text{cms}} = 10 \,\mathrm{MeV}$.
The K-matrix is normalized in such a way that the relationship between the observable phase shift $\delta$ and the on-shell K-matrix has the form $\delta = -\mathrm{arctan}(2 \pi \mu q K)$.

To convert \eqref{eq:lsefork} into a matrix equation \eqref{eq:lse},
we discretize the momenta $p$, $p'$ and $k$ and replace the integration with a discrete sum. For simplicity, we also reduce the upper integration limit from infinity to $p_\text{max}=2$ GeV. This is equivalent to multiplying the potential with an additional sharp momentum-space cutoff.

We tabulate the potential $V (p', \, p)$ and the K-matrix $K (p', \, p)$ on a mesh of initial and final momenta such that $\mathbb{V}$ and $\mathbb{K}$
have a size of $101 \times 101$. 
The first $100$ points of the mesh correspond to Gauss-Legendre quadrature points, rescaled to the interval from 0 to $p_\text{max}$. The last mesh point is the on-shell momentum $q$.
The additional row and column in the matrices $\mathbb{V}$ and $\mathbb{K}$, corresponding to the on-shell momentum $q$, allow us to apply the standard subtraction
method for treating the singularity in the LSE~\eqref{eq:lsefork}.
The element $(101,101)$ of the matrix $\mathbb{K}$ corresponds to the on-shell K-matrix value we are interested in.

After discretization, the LSE~\eqref{eq:lsefork} acquires the matrix from 
$\mathbb{K} = \mathbb{V} + \mathbb{V}\mathbb{G}\mathbb{K}$
with the potential $\mathbb{V}=\mathbb{V}_0 + \mathbb{X}\mathbb{C}\mathbb{Z}$ given in the previous section. Here, $\mathbb{G}$ is a discretized expression for the two-body resolvent operator given by $\mathbb{G} = {\rm diag} (G_1, \ldots , G_{N_{\rm mesh}}, G_{N_{\rm mesh}+1})$ with 
\beq
G_l \; =\;  \frac{2 \mu p_l^2 \Delta p_l}{q^2 - p_l^2} 
\eeq
for $l \leq N_{\rm mesh}$ and 
\beq
G_{N_{\rm mesh}+1} \; =\;  - \sum_{l = 1}^{N_{\rm mesh}} \frac{2 \mu q^2 \Delta p_l}{q^2 - p_l^2} \; +\;  \frac{\mu}{q} \ln \bigg( \frac{p_\text{max}+q}{p_\text{max}-q} \bigg). 
\eeq
Here, $\Delta p_l$ denote the Gauss-Legendre weights and $N_{\rm mesh} = 100$. 
We now build an exact emulator for $\mathbb{K}$. 
To this aim, we apply the LSE Woodbury formula in Eq.~\eqref{eq:woodlse} to the matrix equation $\mathbb{K} = \mathbb{V} + \mathbb{V}\mathbb{G}\mathbb{K}$.
This way, we obtain a semi-analytic expression~\eqref{eq:kmatremu} for the on-shell K-matrix as a function of the parameters $\vec{c}$.
The numerical coefficients $K_0$ and $a_i$ appearing in Eq.~\eqref{eq:kmatremu} are listed in Table~\ref{tab:SM_tab1}.
For any values of the parameters $c_1$, $c_2$, $c_3$, the emulated expression  in 
Eq~\eqref{eq:kmatremu} gives exactly the same result as 
a direct solution of the $101 \times 101$ matrix equation $\mathbb{K} = \mathbb{V} + \mathbb{V}\mathbb{G}\mathbb{K}$, but at a computationally negligible cost. 
\begin{table}[t]
\begin{tabular*}{\textwidth}{@{\extracolsep{\fill}}c@{\extracolsep{\fill}}c@{\extracolsep{\fill}}c@{\extracolsep{\fill}}c@{\extracolsep{\fill}}c@{\extracolsep{\fill}}c@{\extracolsep{\fill}}c@{\extracolsep{\fill}}c@{\extracolsep{\fill}}c@{\extracolsep{\fill}}c@{\extracolsep{\fill}}c}
        \toprule
                $K_0$ & $a_1$ &$a_2$ &$a_3$ &$a_4$ &$a_5$ &$a_6$ &$a_7$ &$a_8$ &$a_9$ &$a_{10}$  \\
        \midrule
$-1.58708$ & $7.74598$ & $-2.72226$ & $-60.58019$ & $-7.74598$ & $-0.03058$ & $1.05010$ & $-2.11053$ & $-12.81704$ & $-1.05010$ & $-0.16881$  \\
        \bottomrule
    \end{tabular*}
       \caption{Pre-calculated numerical coefficients for the emulated K-matrix at $10$~MeV in Eq.~(\ref{eq:kmatremu}). The parameters $K_0$ and $a_{1, \ldots , 5}$ have units of GeV$^{-2}$, while $a_{6, \ldots , 10}$ are dimensionless. Here the values for $K_0$ and $a_i$ are rounded for better readability, unrounded values can be found in the provided Jupyter Notebook. 
   }
    \label{tab:SM_tab1}
\end{table}

\subsubsection{C.~Bound-state problem}
We now switch to the bound-state problem with a constrained binding energy.
We first formulate and discretize the two-body Schr\"odinger equation to cast it into a matrix eigenvector problem. We then use this eigenvector problem in sec.~III.D to illustrate our bound-state emulator.

The bound state we are going to emulate is inspired by the deuteron bound state in  neutron-proton scattering. We consider two particles with a reduced mass $\mu=470\,\text{MeV}$
interacting via the potential $V(p', p, \vec{c}\,)$ defined in Eq.~\eqref{eq:potentialswave}, which form a bound state with a fixed 
energy of $\bar{E} = -2.22\, \text{MeV}$.
The potential in Eq.~\eqref{eq:potentialswave} depends on three adjustable parameters 
$c_1$, $c_2$, and $c_3$, but the requirement of the eigenenergy to be exactly $\bar{E} = -2.22\,\text{MeV}$ places a constraint on possible values of these parameters. 
The bound-state wave function $\Psi^{\bar E}(p)$  
can be obtained by solving the S-wave bound-state Schr\"odinger equation in momentum space,
\begin{equation}
    \frac{p^2}{2\mu} \Psi^{\bar E}(p) + \int_{0}^{\infty}V(p', p, \vec{c}\,) \Psi^{\bar E}(p') p'^2 dp' = \bar{E} \Psi^{\bar E}(p),
    \label{eq:scroed}
\end{equation}
and is normalized such that $\int_{0}^{\infty} \Psi^{\bar E}(p)^2 p^2 dp = 1$.
The corresponding coordinate-space wave function $u^{\bar E}(r)$ can be computed from $\Psi^{\bar E}(p)$ by means of a three-dimensional Fourier transform, which for the S-wave component reduces to the spherical Bessel transform
\begin{equation}
    \frac{u^{\bar E}(r)}{r} = \sqrt{\frac{2}{\pi}} \int_0^{\infty} \Psi^{\bar E}(p) j_0(p r) p^2 dp,
    \label{eq:fouriertr}
\end{equation}
where $j_0(x)$ denotes a spherical Bessel function.
As an example of bound-state observable, we consider the mean-square (m.s.) matter radius $r_m^2$, defined as
\begin{equation}
    r_m^2 = \frac{\int_{0}^{\infty} {u^{\bar E}(r)}^2 \frac{r^2}{4} dr}{\int_{0}^{\infty} {u^{\bar E}(r)}^2 dr}.
    \label{eq:matterradius}
\end{equation}
We will use this setup to benchmark our emulator.

As a next step, we discretize the equations~\eqref{eq:scroed}-\eqref{eq:matterradius}.
We replace the upper integration limits in Eqs.~\eqref{eq:scroed}-\eqref{eq:fouriertr} with $p_{\max} = 2.0$ GeV and in Eq.~\eqref{eq:matterradius} with  $r_{\max} = 50$ fm.
For the momenta $p$ and $p'$, we use  $100$ Gauss-Legendre meshpoints $p_i$ and weights $\Delta p_i$, rescaled to the interval from $0$ to $p_{\max}$.
For the coordinate $r$, we use $1000$ Gauss-Legendre meshpoints $r_i$ and weights $\Delta r_i$, rescaled to the interval from $0$ to $r_{\max}$.

Next, we discretize the wave function ($[\Psi^{\bar E}]_i \equiv \Psi^{\bar E}(p_i)$)  and the potential ($[\mathbb{V}]_{ij} \equiv V(p_i, p_j)$) and replace the intergal in Eq.~\eqref{eq:scroed} with a sum using the Gauss-Legendre quadrature:
\begin{equation}
     \sum_j 
     \bigg(
     \delta_{ij} \frac{p_j^2}{2\mu}  + [\mathbb{V}]_{ij}  p_j^2 \Delta p_j 
     \bigg) [\Psi^{\bar E}]_j
     = \bar E [\Psi^{\bar E}]_i.
    \label{eq:scroeddiscr}
\end{equation}
Now, we introduce the Hamiltonian matrix $\mathbb{H}$ via
\begin{equation}
    [\mathbb{H}]_{ij} \equiv \delta_{ij} \frac{p_j^2}{2\mu}  + [\mathbb{V}]_{ij}  p_j^2 \Delta p_j,
    \label{eq:hamiltdiscr}
\end{equation}
such that~\eqref{eq:scroed} takes the form of a standard matrix eigenvalue problem:
\begin{equation}
    \mathbb{H} \Psi^{\bar E} = \bar E \Psi^{\bar E}.
    \label{eq:scroeddiscrshort}
\end{equation}
The spherical Bessel transformation~\eqref{eq:fouriertr} is a linear transformation and for discretized functions can be rewritten as a matrix-vector product
\begin{equation}
    u^{\bar E} = \mathbb{F} \Psi^{\bar E},
\end{equation}
where $[u^{\bar E}]_j \equiv u^{\bar E}(r_j)$. Notice that due to the highly oscillating behavior of the spherical Bessel function for large argument values, it is advantageous to employ a  denser integration grid in Eq.~\eqref{eq:fouriertr} using interpolation to reconstruct the smooth function $\Psi^{\bar E} (p)$ from $\Psi^{\bar E} (p_i)$. The resulting numerical implementation of the matrix $\mathbb{F}$ can be found in the provided Jupyter Notebook.  

Finally, to discretize the expression for the m.s.~matter radius in  Eq.~\eqref{eq:matterradius}, we approximate both integrals using the Gauss-Legendre quadrature and introduce two auxiliary diagonal matrices $[\mathbb{r}]_{ii} \equiv  r_i^2 \Delta r_i / 4$ and $[\mathbb{n}]_{ii} \equiv  \Delta r_i $. This leads to the discretized version of Eq.~\eqref{eq:matterradius}:
\begin{equation}
    r_m^2 = 
    \frac{(u^{\bar E})^\dagger \mathbb{r} \, u^{\bar E}}{(u^{\bar E})^\dagger \mathbb{n} \, u^{\bar E}} = 
    \frac{(\Psi^{\bar E})^\dagger \mathbb{F}^\dagger \mathbb{r} \, \mathbb{F} \Psi^{\bar E}}{(\Psi^{\bar E})^\dagger \mathbb{F}^\dagger \mathbb{n} \, \mathbb{F} \Psi^{\bar E}}
    .
    \label{eq:matterradiusdiscr}
\end{equation}
In the following section, we will build an exact emulator for the fixed-energy bound-state problem~\eqref{eq:scroeddiscrshort} and show how to emulate observables such as, e.g., the  m.s.~matter radius in Eq.~\eqref{eq:matterradiusdiscr}.

\subsubsection{D.~Emulation of the bound-state wave function and observables}
\label{subsec:exampleWFemul}

We now build the exact emulator for a fixed-energy wave function for the considered interaction model. To this aim, we define a fixed-eigenvalue problem,
project it on a low-dimensional subspace, find a condition on the parameters that keeps the eigenvalue fixed, and finally obtain the emulator for the wave function corresponding to a fixed eigenenergy. We also present an emulator for the mean-square matter radius as a representative example of an observable which is directly computable from the eigenvector.

We consider the following matrix eigenvalue problem (formulated in Eq.~\eqref{eq:scroeddiscrshort} of the previous section)
\begin{equation}
    \left( \mathbb{H}_0 + c_1 \mathbb{H}_1 + c_2 \mathbb{H}_2 + c_3 \mathbb{H}_3  \right) \Psi^{\bar E} = \bar E \Psi^{\bar E},
    \label{eq:scroeddiscrshortwithparams}
\end{equation}
where $\bar{E}$ is fixed to the value $\bar{E} = -2.22\, \text{MeV}$.
The parameters $c_1$, $c_2$, $c_3$ can take arbitrary values, subject to the constraint that $\bar{E} = -2.22\, \text{MeV}$. The matrices $\mathbb{H}_i$ are known, and their definition can be extracted from Eqs.~\eqref{eq:hamiltdiscr} and~\eqref{eq:toypotentialmatrix}.
Our goal is to accurately and efficiently compute the eigenvector $\Psi^{\bar E}$ for any values of $c_1$, $c_2$, $c_3$ that keep the eigenenergy fixed at $\bar{E} = -2.22\, \text{MeV}$.

As discussed in the previous section, the dimension of matrices $\mathbb{H}_i$ is $100 \times 100$. Moreover, the matrices $\mathbb{H}_{1,2,3}$ of the parameter-dependent updates have low ranks. Specifically, $\mathbb{H}_1$ and $\mathbb{H}_3$ are of rank $1$, while $\mathbb{H}_2$ has rank $2$. The combined rank of parameter-dependent terms is thus at most 4.
We can always rewrite the full Hamiltonian matrix $\mathbb{H}_0 + c_1 \mathbb{H}_1 + c_2 \mathbb{H}_2 + c_3 \mathbb{H}_3$ in the form $\mathbb{H}_0 + \mathbb{X}\mathbb{C}\mathbb{Z}$, where all parameters $c_i$ only appear in the matrix $\mathbb{C}$. Since the combined rank of update does not exceed 4,
the dimensions of matrix $\mathbb{C}$ should also not exceed $4 \times 4$.

The fixed-eigenvalue problem in Eq.~\eqref{eq:scroeddiscrshortwithparams} satisfies the conditions discussed in sec.~II. The dimensionality of the subspace spanned by the vectors $\Psi^{\bar E} (\vec c \, )\big|_{E(\vec c \, ) = \bar E}$, which fulfill Eq.~\eqref{eq:scroeddiscrshortwithparams}, is therefore limited by 4.
Consequently, any eigenvector $\Psi^{\bar E}$ that satisfies Eq.~\eqref{eq:scroeddiscrshortwithparams} can be expressed as a linear combination of at most 4 basis vectors. Once we find suitable basis vectors, we can project Eq.~\eqref{eq:scroeddiscrshortwithparams} on this reduced basis.
Such a projection does not result in accuracy loss, since the exact solution of the original Eq.~\eqref{eq:scroeddiscrshortwithparams} itself spans a very low-dimensional subspace. The appropriately chosen reduced basis can, therefore, be used to generate exact solutions of the original (not projected) problem in Eq.~\eqref{eq:scroeddiscrshortwithparams}.

To find basis vectors, which can optimally represent any valid solution $\Psi^{\bar E}$, we need to find several (at most 4) linearly independent solutions (snapshots) of Eq.~\eqref{eq:scroeddiscrshortwithparams}
and orthonormallize them. 
We proceed as follows: we first choose several fixed values for the parameters $c_2$ and $c_3$ and then fit $c_1$ to satisfy the equation
\begin{equation}
    \det(\mathbb{H}_0 + c_1 \mathbb{H}_1 + c_2 \mathbb{H}_2 + c_3 \mathbb{H}_3 - \bar{E} \, \mathbb{1})=0 .
\end{equation}
In this way, we find the following combinations of parameters $\{ c_1, c_2, c_3\}$, which lead to the desired binding energy $\bar E$: $\{-0.0673, -0.1, -0.1 \}$, $\{-0.0976, 0.1, -0.1 \}$, $\{-0.0687, -0.1, 0.1 \}$, $\{-0.0984, 0.1, 0.1 \}$. Here and in what follows, the various precomputed numerical coefficients are rounded for better readability, unrounded values can be found in the provided Jupyter Notebook.
Next, we compute 4 direct solutions (snapshots) of Eq.~\eqref{eq:scroeddiscrshortwithparams} using these sets of parameters.
We combine four resulting eigenvectors in a single $4 \times 100$ matrix and perform SVD of it. We observe that there are just two non-zero singular values.
We choose the right singular vectors corresponding to the non-zero singular values as our reduced basis.
We denote a $100 \times 2$ basis matrix constructed from these two vectors as $\mathbb{B}$.
Any solution of Eq.~\eqref{eq:scroeddiscrshortwithparams} can be represented as
\begin{equation}
    \Psi^{\bar E} = \mathbb{B} \Psi^{\bar E}_\text{rb},   
    \label{eq:wfrb}
\end{equation}
where $\Psi^{\bar E}_\text{rb}$ denotes the wave function in the reduced basis (i.e., $\Psi^{\bar E}_\text{rb}$ is a 2-dimensional vector).

We can now project the eigenvalue problem Eq.~\eqref{eq:scroeddiscrshortwithparams} onto the reduced basis $\mathbb{B}$. To this aim, we substitute Eq.~\eqref{eq:wfrb} into Eq.~\eqref{eq:scroeddiscrshortwithparams} and multiply the result by $\mathbb{B}^\dagger$ from the left to obtain
\begin{equation}
    \left( \mathbb{H}_{0\text{rb}} + c_1 \mathbb{H}_{1\text{rb}} + c_2 \mathbb{H}_{2\text{rb}} + c_3 \mathbb{H}_{3\text{rb}}  \right) \Psi^{\bar E}_{\text{rb}} = \bar E \Psi^{\bar E}_{\text{rb}},
    \label{eq:scroeddiscrshortwithparamsRB}
\end{equation}
where $\mathbb{H}_{i\text{rb}} \equiv \mathbb{B}^\dagger \mathbb{H}_{i} \mathbb{B}$ are the reduced-basis $2 \times 2$ Hamiltonian matrices that can be pre-computed. 

Next, we use the projected equation \eqref{eq:scroeddiscrshortwithparamsRB} to find the condition on $\vec c\, $ to ensure that the eigenenergy is exactly fixed at the value $\bar E$. This condition corresponds to a solution of the equation 
\begin{equation}
    \det( \mathbb{H}_{0\text{rb}} + c_1 \mathbb{H}_{1\text{rb}} + c_2 \mathbb{H}_{2\text{rb}} + c_3 \mathbb{H}_{3\text{rb}}  - \bar{E} \, \mathbb{1}_{})=0.
\end{equation}
Using the pre-computed $2 \times 2$ matrices $\mathbb{H}_{i\text{rb}}$, we solve this equation symbolically to obtain $c_1$ as a function of $c_2$ and $c_3$. For the case at hand, we find:
\begin{equation}
c_1=\frac{- 0.54882 c_{2}^{2} + 0.94125 c_{2} + 0.081345 c_{3} + 0.52622}{- 0.54882 c_{3} - 6.2731}\,.
\label{eq:exampleConstraint}
\end{equation}
Equation~\eqref{eq:exampleConstraint} provides the constraint on the parameters 
$\vec c\,$  that has been implicitly imposed by fixing the eigenenergy to $\bar{E} = -2.22\, \text{MeV}$ in Eq.~\eqref{eq:scroeddiscrshortwithparams}.

The reduced eigenvalue problem \eqref{eq:scroeddiscrshortwithparamsRB} has a dimension of just $2 \times 2$ and can be solved symbolically, leading to a symbolic expression for (non-normalized) $\Psi^{\bar E}_{\text{rb}}$ : 
\begin{equation}
 \Psi^{\bar E}_{\text{rb}} = \begin{bmatrix}
                               - 0.81823 c_{1} - 0.16027 c_{2} - 0.00045886 c_{3} + 0.010909  \\
                                - 0.026241 c_{1} - 0.0020622 c_{2} - 5.7718 \cdot 10^{-6} c_{3} - 0.002187
                              \end{bmatrix}.
\end{equation}
Using Eq.~\eqref{eq:wfrb}, this then allows one to reconstruct the original eigenvector $\Psi^{\bar E}$ as a function of $\vec c$.  
Finally, once we have the emulator for the eigenvector $ \Psi^{\bar E}$, we can also emulate bound-state observables. For example, the emulated expression for the m.s.~matter radius in Eq.~\eqref{eq:matterradiusdiscr} has the form
\begin{equation}
r_m^2 = \frac{
b_1 c_1^2 + b_2 c_1 c_2 + b_3 c_1 c_3 + b_4 c_1
+ b_5 c_2^2 + b_6 c_2 c_3 + b_7 c_2
+ b_8 c_3^2 + b_9 c_3 + b_{10}
}{
d_1 c_1^2 + d_2 c_1 c_2 + d_3 c_1 c_3 + d_4 c_1
+ d_5 c_2^2 + d_6 c_2 c_3 + d_7 c_2
+ d_8 c_3^2 + d_9 c_3 + d_{10}
},
\label{eq:radius}
\end{equation}
with the pre-computed numerical coefficients $b_i$ and $d_i$ listed in Table~\ref{tab:radiusparams}. It is important to emphasize that Eq.~\eqref{eq:radius} is only valid for values of the parameters $c_1$, $c_2$ and $c_3$, which satisfy the condition in Eq.~\eqref{eq:exampleConstraint}.
Last but not least, we emphasize, once again, that the considered emulator does not incur any accuracy loss.

\begin{table}[t]
\begin{tabular*}{\textwidth}{@{\extracolsep{\fill}}l@{\extracolsep{\fill}}c@{\extracolsep{\fill}}c@{\extracolsep{\fill}}c@{\extracolsep{\fill}}c@{\extracolsep{\fill}}c@{\extracolsep{\fill}}c@{\extracolsep{\fill}}c@{\extracolsep{\fill}}c@{\extracolsep{\fill}}c@{\extracolsep{\fill}}c}
        \toprule
       $i$    &     $1$ & $2$ &$3$ &$4$ &$5$ &$6$ &$7$ &$8$ &$9$ &${10}$   \\
        \midrule
$b_i$
 & $ 38696 $ & $ 14904 $ & $ 42.661 $ & $ -821.8 $ & $ 1439.2 $ & $ 8.2397 $ & $ -165.14 $ & $ 0.011793 $ & $ -0.47299 $ & $ 7.1978 $\\
$d_i$  & $ 360.17 $ & $ 138.17 $ & $ 0.39547 $ & $ -7.1921 $ & $ 14.669 $ & $ 0.0841 $ & $ -3.7187 $ & $ 0.00012054 $ & $ -0.010747 $ & $ 1  $\\
        \bottomrule
    \end{tabular*}
       \caption{Pre-computed coefficients in the emulated expression for the mean-squared matter radius in Eq.~(\ref{eq:radius}). The coefficients $b_i$ are given in units of GeV$^{-2}$, while $d_i$ are dimensionless.
   }
    \label{tab:radiusparams}
\end{table}

\subsection{IV.~Implementation setup for nucleon-deuteron scattering}

We now briefly comment on our calculational setup for the nucleon-deuteron scattering emulator. As the interaction model, we use the SMS \cite{Reinert:2017usi} 2N potential at N$^4$LO$^+$ corresponding to the  cutoff value of $\Lambda=450\,\text{MeV}$ and the three-nucleon force at N$^2$LO specified in Ref.~\cite{Maris:2020qne}.
To solve the Faddeev equation, see Ref.~\cite{Gloeckle:1995jg} for details, we use 37 $q$-points and 32 $p$-points for discretization while keeping the energy fixed to the value of $E_{N,\text{lab}}=70\,\text{MeV}$. For the partial wave expansion, we truncate the two-body total angular momentum to the value of $j_{\text{max}}=5$. We present our results for the channel with $J^\pi=\frac12^+$, to which both the $c_D$- and $c_E$-parts of the three-nucleon force contribute. Finally, for the emulator of $T^{3N} (c_E)$, we keep $c_D$ at the value of $c_D = 0.8891$ obtained for this interaction model using the standard LENPIC fitting protocol as explained in Ref.~\cite{Maris:2020qne}.

\end{document}